\documentclass[11pt,a4paper]{article}
\usepackage{jheppub}
\usepackage[T1]{fontenc}
\usepackage{lmodern}
\usepackage{mathtools,booktabs,array,microtype}

\newcommand{\Res}{\operatorname{Res}}
\newcommand{\PT}{\operatorname{PT}}
\newcommand{\Nfour}{\mathcal N=4}
\newcommand{\A}{\mathcal A}
\newcommand{\id}{\mathbf 1}
\newcommand{\ee}{\epsilon}

\makeatletter
\gdef\@fpheader{\vbox{\small\hfill \par \vskip 14pt}}
\makeatother

\title{All-multiplicity completion of maximally supersymmetric gauge amplitudes}
\author{Ilmo Sung}
\affiliation{Science and Technology Directorate, U.S. Department of Homeland Security, \\
Washington, DC, USA}
\emailAdd{ilmo.sung@hq.dhs.gov}
\abstract{
We show that the known six-point constraints on four-point amplitudes are sufficient for completion at arbitrary multiplicity in planar, four-dimensional maximally supersymmetric gauge theory. We construct massless tree amplitudes as formal series in the derivative expansion, with the standard supersymmetric Yang-Mills three-point amplitudes, unbroken R-symmetry, and parity invariance of amplitudes with only external scalars, holding the scalar labels fixed. For a positive leading four-point contact coefficient, the even coefficients in the logarithmic parameterization are fixed, while every real sequence of odd coefficients is realized by a completion at all derivative orders. The construction uses the field-theory Kawai-Lewellen-Tye bilinear and normalized matrices of sphere integrals. Exponentiating homogeneous components of their matrix logarithms varies the odd coefficients independently while preserving locality and factorization through the complete massless multiplet. The six-point conditions are therefore necessary and sufficient for such a completion. Higher-multiplicity consistency imposes no further restrictions on these four-point coefficients once the higher-point contact interactions are chosen consistently with the same symmetries and factorization conditions.
}
\keywords{Scattering Amplitudes, Supersymmetric Gauge Theory,\\ Effective Field Theories}
\begin{document}
\maketitle
\flushbottom

\section{Introduction}\label{sec:intro}

Every four-point amplitude satisfying the known six-point constraints admits an extension to local massless tree amplitudes at arbitrary multiplicity in the class considered here. We work in four-dimensional planar maximally supersymmetric gauge theory, retain the standard super Yang-Mills (SYM) three-point amplitudes and unbroken $SU(4)$ R-symmetry, and impose parity invariance on amplitudes with only external scalars, holding their labels fixed. For a positive leading four-point contact coefficient, the construction realizes every allowed real coefficient sequence at all orders in the derivative expansion. Higher-multiplicity consistency therefore imposes no additional restriction on these four-point coefficients when the higher-point contact interactions are chosen to satisfy the same symmetries and factorization conditions.

At six points, a three-particle factorization channel contains a product of two four-point amplitudes. Compatibility with supersymmetry and the scalar parity condition therefore imposes nonlinear relations on four-point data~\cite{Sung2026}. At higher multiplicity, the factorization equations also involve additional lower-point amplitudes and contact interactions. Independent local interactions are the data not fixed by lower-point residues in general on-shell constructions~\cite{CohenElvangKiermaier2011}. The completion problem is to choose these interactions so that every residue contains the corresponding lower-point members of one amplitude family.

The preceding analysis~\cite{Sung2026} reconstructed the angular dependence of the four-point amplitude from its regular forward limit, expressing its normalized multiplier through the exponential of a one-variable function. The stronger six-point condition derived subsequently in ref.~\cite{HWWW2026} fixes the even coefficients of this function in terms of its quadratic coefficient and leaves the odd coefficients arbitrary. These relations characterize the necessary four-point data to all derivative orders. The associated coefficient freedom was identified in earlier supersymmetric calculations~\cite{EHM2026} and biadjoint scalar effective theories~\cite{CEH2022,CEH2023}. Reference~\cite{EHM2026} anticipated the absence of further algebraic relations among the odd coefficients using conjectural algebraic independence of odd zeta values. Here we establish the stronger existence statement: every real assignment of the complete sequence is realized by a single compatible amplitude family at all multiplicities and derivative orders.

The construction uses the field-theory Kawai-Lewellen-Tye (KLT) bilinear. The disk integrals reproduce the required even four-point coefficients. We expand the matrix logarithm of the normalized matrix of sphere integrals and show that exponentiating each homogeneous component defines scalar amplitudes that satisfy locality and factorization at every multiplicity. Components of odd degree vary the odd coefficients of the four-point logarithm independently. Their exponentials determine the higher-order terms required by nonlinear factorization. A final KLT bilinear with SYM amplitudes preserves the complete massless multiplet and the supersymmetry Ward identities.

Completeness of the four-point conditions does not determine all higher-point interactions. Using the commutator of the coefficient matrices already present in the string-amplitude expansion~\cite{SchlottererStieberger2012}, we construct a real one-parameter deformation that preserves the entire four-point amplitude and extends to every higher multiplicity. Its nonzero five-point contact amplitude distinguishes the resulting completions for every fixed choice of four-point data.

Section~\ref{sec:data} specifies the amplitude class and derives the four-point amplitudes to be completed from the known six-point identities. Section~\ref{sec:klt} introduces the scalar amplitudes and establishes factorization of their KLT composition. Section~\ref{sec:construction} constructs the amplitudes for arbitrary coefficients and verifies their physical properties. Section~\ref{sec:examples} gives examples and the five-point deformation, and section~\ref{sec:discussion} discusses the resulting restrictions on four-point amplitudes. The appendices give integral conventions, details of the ordering bases and formal series, and the explicit five-point calculation.

\section{Four-point amplitudes and the completion problem}\label{sec:data}

\subsection{Amplitude class and conventions}

We consider planar color-ordered tree amplitudes with the four-dimensional massless $\Nfour$ SYM multiplet. All momenta are outgoing, the coupling $g$ is real and nonzero, and every three-point amplitude is the standard SYM amplitude. Complexified momenta define three-point amplitudes and generic massless factorization channels.

For a reference mass $\Lambda>0$, introduce dimensionless invariants
\begin{equation}
 s_I=\frac{\bigl(\sum_{i\in I}p_i\bigr)^2}{\Lambda^2},
 \qquad s=s_{12},\quad u=s_{14},\quad t=s_{13},
 \qquad s+t+u=0.
 \label{eq:kinematics}
\end{equation}
The spinors are expressed in units of the reference scale $\Lambda$. We use the bracket convention of ref.~\cite{Sung2026},
\begin{equation}
 p_i/\Lambda=\lambda_i\tilde\lambda_i^{\mathsf T},\qquad
 \langle ij\rangle=\det(\lambda_i,\lambda_j),\qquad
 [ij]=\det(\tilde\lambda_i,\tilde\lambda_j),\qquad
 s_{ij}=\langle ij\rangle[ij].
 \label{eq:spinors}
\end{equation}
Component amplitudes are expressed in the same units, $\A_n=\Lambda^{n-4}\A_{n,\mathrm{phys}}$. We use phase conventions in which the propagator is $+1/s_I$, with the signs of the three-point amplitudes and the phases of crossed states fixed consistently. Restoring dimensions gives $1/P_I^2$.

The amplitudes are formal derivative expansions. A variable $\ee$ records degree in Mandelstam invariants after the common leading homogeneous scaling is removed; one power of $\ee$ corresponds to two additional powers of momenta. At each degree, locality requires polynomial contact terms and simple massless poles in compatible planar channels. Compatible simultaneous poles are those of a common planar tree; crossing diagonals of its ordering polygon cannot occur together. The complete series is treated formally, with no convergence assumption.

The amplitudes obey Lorentz and gauge invariance, all sixteen supersymmetry Ward identities, unbroken $SU(4)$ R-symmetry, reality, and covariance under relabeling. Cyclic orderings retain their orientation, with the standard reflection signs. Every amplitude with only external scalars satisfies
\begin{equation}
 \A_n(\phi_{a_1},\ldots,\phi_{a_n};\lambda,\tilde\lambda)
 =\A_n(\phi_{a_1},\ldots,\phi_{a_n};\tilde\lambda,\lambda).
 \label{eq:scalarparity}
\end{equation}
The scalar labels $a_i$ remain fixed. On complexified kinematics the condition is an algebraic exchange of spinors; on real kinematics it imposes parity invariance on the specified scalar amplitude.

For a physical channel $q=s_I$, let $\Res_{q=0}$ denote the coefficient of $1/q$. Factorization requires
\begin{equation}
 \Res_{q=0}\A_n
 =\sum_{h\in\mathcal H_{\rm SYM}}
 \A_L(\ldots,P_I^h)\,
 \A_R((-P_I)^{\bar h},\ldots).
 \label{eq:physicalfactorization}
\end{equation}
Here $P_I$ is outgoing on the left, and $\bar h$ is the crossed state on the right. The sum uses the standard positive inner product on all sixteen physical SYM states. The lower amplitudes are members of the same family, and compatible iterated residues agree. Higher-point contact coefficients are allowed to vary subject to these requirements; a completion consists of a compatible choice of those coefficients.

\subsection{The four-point family}

At four points the supersymmetry Ward identities~\cite{EFK2010} permit a scalar multiplier,
\begin{equation}
 \A_4=B(s,u)\A_4^{\rm SYM},\qquad
 F(s,u)=-\frac{g^2}{su}B(s,u)
       =-\frac{g^2}{su}+f(s,u),
 \label{eq:fourpointnormalization}
\end{equation}
where $f$ is a regular formal series. The four-gluon component with negative helicities on legs 1 and 2 is $\langle12\rangle^2[34]^2F(s,u)$.\footnote{In physical units, $F_{\rm phys}(\Lambda^2s,\Lambda^2u)=\Lambda^{-4}F(s,u)$. The multiplier denoted $M$ in ref.~\cite{Sung2026} satisfies $B(s,u)=M(\Lambda^2s,\Lambda^2u)$.}

The six-point reconstruction relation is~\cite{Sung2026}
\begin{equation}
 B(x,y)B(x+y,z)=B(x,y+z)B(y,z),
 \label{eq:cocycle}
\end{equation}
with $B(s,u)=B(u,s)$ and $B(s,0)=B(0,u)=1$. To obtain its solution, take $b=\log B$ and differentiate with respect to $z$ at $z=0$:
\begin{equation}
 \partial_u b(s,u)=\partial_2 b(s+u,0)-\partial_2 b(u,0).
 \label{eq:cocyclederivative}
\end{equation}
Here $\partial_2$ differentiates the second argument. Define $H'(s)=-\partial_2 b(s,0)$ and $H(0)=0$. Integration in $u$ then gives
\begin{equation}
 B(s,u)=\exp\!\bigl[H(s)+H(u)-H(s+u)\bigr],\qquad
 H(s)=\sum_{k\ge2}\frac{c_k}{k}s^k.
 \label{eq:generator}
\end{equation}
A linear term in $H$ cancels from $B$; this definition sets it to zero. The coefficients have a direct forward-amplitude interpretation. Since $B(s,u)=1-su\,f(s,u)/g^2$, differentiation with respect to $u$ at $u=0$ gives
\begin{equation}
 H'(s)=\frac{s}{g^2}f(s,0),\qquad
 f(s,0)=g^2\sum_{k\ge2}c_k s^{k-2}.
 \label{eq:Hforward}
\end{equation}
Thus $H$ determines the regular forward amplitude. Its odd-indexed coefficients $c_{2j+1}$ are coefficients of this one-variable function, not independently assigned Taylor coefficients of the full two-variable amplitude.\footnote{The coefficients $a_{n,0}$ of ref.~\cite{HWWW2026} are the forward coefficients of $f/g^2$ in the present normalization, so $a_{n,0}=c_{n+2}$.}

The scalar parity condition at six points also implies the three-term identity~\cite{HWWW2026}
\begin{equation}
 \begin{aligned}
 &(x_1x_3-x_2x_4)B(x_1,x_2)B(x_3,x_4)\\
 &\quad+x_2x_4\,B(x_1,x_2+x_3)B(x_3,x_1+x_4)\\
 &=x_1x_3\,B(x_2,x_1+x_4)B(x_2+x_3,x_4),
 \end{aligned}
 \label{eq:sixpointidentity}
\end{equation}
where $x_1,x_2,x_3$ are independent formal variables and $x_4=-x_1-x_2-x_3$. This is eq.~(45) of ref.~\cite{HWWW2026} in the normalization~\eqref{eq:fourpointnormalization}. Its derivation uses the massless spectrum, cubic amplitudes, supersymmetry, R-symmetry, and scalar parity specified above; it does not impose a massive spectrum or dispersive assumptions.

To determine the even part in the exponential representation~\eqref{eq:generator}, introduce an auxiliary function as in ref.~\cite{HWWW2026}, section~3.1:
\begin{equation}
 H_{\mathrm{even}}(x)=\frac{H(x)+H(-x)}2,
 \qquad
 \varphi(x)=x\,e^{-2H_{\mathrm{even}}(x)}.
 \label{eq:evenauxiliary}
\end{equation}
The series $\varphi$ is odd and begins as $\varphi(x)=x-c_2x^3+O(x^5)$. With $x_1=x$, $x_2=y$ and $x_3=z$, substitution into eq.~\eqref{eq:sixpointidentity} and cancellation of a common exponential with constant term one give
\begin{equation}
 \varphi(x)\varphi(z)+\varphi(y)\varphi(x+y+z)
 =\varphi(x+y)\varphi(y+z).
 \label{eq:evenadditionidentity}
\end{equation}
The odd part of $H$ cancels. Differentiating with respect to $y$ and setting $y=0$ yields
\begin{equation}
 \varphi(x+z)=\varphi'(x)\varphi(z)+\varphi(x)\varphi'(z).
 \label{eq:evenadditionlaw}
\end{equation}
Differentiating twice with respect to $z$ at $z=0$, using oddness and $\varphi'''(0)=-6c_2$, gives
\begin{equation}
 \varphi''(x)=-6c_2\varphi(x),
 \qquad \varphi(0)=0,\qquad \varphi'(0)=1.
 \label{eq:evenODE}
\end{equation}
For $c_2>0$, set $\kappa=\sqrt{6c_2}$. The unique normalized formal solution is
\begin{equation}
 \varphi(x)=\frac{\sin(\kappa x)}{\kappa},
 \qquad
 H_{\mathrm{even}}(x)=\frac12\log\frac{\kappa x}{\sin(\kappa x)}.
 \label{eq:evensolution}
\end{equation}
The quotient $\sin(\kappa x)/(\kappa x)$ has constant term one, so the logarithm is defined as a formal series. Conversely, the sine solution satisfies the full identity~\eqref{eq:evenadditionidentity}. Reversing the exponential cancellation shows that this even part, together with any real formal odd part of $H$, satisfies eq.~\eqref{eq:sixpointidentity}. Thus the functional identities constrain only $H_{\mathrm{even}}$ once the cocycle representation has been imposed.

Expanding eq.~\eqref{eq:evensolution} gives the four-point amplitudes to be completed. The complete real formal solution in the $c_2>0$ sector is
\begin{align}
 B(s,u)&=\exp\!\left[\sum_{k\ge2}\frac{c_k}{k}P_k(s,u)\right],
 &P_k(s,u)&=s^k+u^k-(s+u)^k,
 \label{eq:Bfamily}\\
 c_2&>0,
 &c_{2j}&=-\frac{B_{2j}}{2(2j)!}(-24c_2)^j,
 \qquad j\ge1,
 \label{eq:evenfamily}
\end{align}
where $B_{2j}$ are Bernoulli numbers with $B_2=1/6$, and each $c_{2j+1}$, $j\ge1$, is an arbitrary real number. For example,
\begin{equation}
 c_4=\frac25c_2^2,\qquad c_6=\frac8{35}c_2^3.
 \label{eq:forward}
\end{equation}
Thus $c_2>0$ specifies a positive leading four-point contact coefficient, $f(0,0)=g^2c_2$. It is a condition on this coefficient, rather than an assumption of spectral positivity. The full real sequence $(c_3,c_5,\ldots)$ is unrestricted in growth. The remaining question is whether every member of this family admits amplitudes at all higher multiplicities.

\subsection{Completion at every multiplicity}

For every $g\ne0$, $c_2>0$, and arbitrary real sequence $(c_3,c_5,\ldots)$ in eqs.~\eqref{eq:Bfamily} and \eqref{eq:evenfamily}, there exists an amplitude family satisfying the conditions of section~\ref{sec:data}, with
\begin{equation}
 \A_3=\A_3^{\rm SYM},\qquad
 \A_4=B(s,u)\A_4^{\rm SYM}.
 \label{eq:result}
\end{equation}
The family is defined for every $n\ge3$ and every derivative order. Its truncations agree as the order is increased, and its lower-point members appear in every factorization channel. Sections~\ref{sec:klt} and \ref{sec:construction} establish this sufficiency statement by construction. Necessity follows from the six-point identities~\eqref{eq:cocycle} and \eqref{eq:sixpointidentity}, whose complete formal solution in this sector was derived above. Combining the two implications makes eqs.~\eqref{eq:Bfamily} and \eqref{eq:evenfamily} necessary and sufficient for four-point data to admit completion in the stated $c_2>0$ class.

\section{KLT composition and factorization}\label{sec:klt}

\subsection{Double-partial amplitudes and the KLT kernel}\label{sec:scalarbasis}

Let $m_n(\alpha\mid\beta)$ be the tree amplitude of a massless cubic biadjoint scalar theory with two cyclic orderings $\alpha,\beta$. The two ordering indices label its two color structures, not distinct physical polarization multiplets. The amplitude is the signed sum over cubic trees planar in both orderings, with a propagator $1/s_I$ for each internal edge. The signs follow from the antisymmetric cubic color vertices. We normalize $m_3(123\mid123)=1$; reversing either ordering gives $m_3(123\mid132)=-1$. At $n$ points, reflection of either cyclic ordering gives the sign $(-1)^n$.

The Kleiss-Kuijf (KK) and Bern-Carrasco-Johansson (BCJ) relations reduce an ordering argument to $(n-3)!$ independent amplitudes~\cite{KK1989,BCJ2008,BBDV2009}. For $n\ge4$, choose the complementary bases
\begin{equation}
 (1,\sigma(2,\ldots,n-2),n-1,n),\qquad
 (1,\tau(2,\ldots,n-2),n,n-1).
 \label{eq:bases}
\end{equation}
Here $\sigma$ and $\tau$ run over all permutations of $2,\ldots,n-2$. The matrix $m_n$, with rows in the first basis and columns in the second, has size $(n-3)!\times(n-3)!$ and is generically invertible. Its inverse
\begin{equation}
 S_n=m_n^{-1}
 \label{eq:kernel}
\end{equation}
is the polynomial field-theory KLT momentum kernel~\cite{KLT1986,BBDF2010,BBDSV2011}. The identification with the inverse biadjoint matrix follows from the scattering-equation representation~\cite{CHY2013}. It has degree $n-3$ in Mandelstam invariants, with signs fixed by the same cubic and propagator conventions. At three points we use the reference ordering $123$ for both basis elements, so $m_3=S_3=1$.

For amplitude families $X,Y$ satisfying field-theory KK and BCJ relations in the ordering indices being contracted, define
\begin{equation}
 (X\star Y)_n=X_nS_nY_n.
 \label{eq:composition}
\end{equation}
The contracted ordering indices are the right index of $X_n$ and the left index of $Y_n$. Any uncontracted indices label the external ordering or the external states. For scalar families satisfying the relations in both ordering arguments, composition is associative and $m$ is its identity, as in the scalar formulation of the KLT bootstrap~\cite{ChiEtAl2022,CEH2023}.

When $X_n$ has leading term $m_n$, multiplication by $S_n$ gives a matrix with identity leading term because $m_nS_n=\id$. Define
\begin{equation}
 R_{X,n}=X_nS_n=\id+O(\ee),\qquad
 R_{X\star Y,n}=R_{X,n}R_{Y,n}.
 \label{eq:Rnormalization}
\end{equation}
The normalized matrices act on the first ordering space. At four points they are scalars; at five points they are $2\times2$ matrices. Thus four-point logarithms add under composition, whereas higher-point matrices can fail to commute. Appendix~\ref{app:bases} gives their transformations under a change of basis.

Higher-derivative double-copy constructions include amplitudes with cubic and quartic field-strength interactions~\cite{BroedelDixon2012,Garozzo2019}, scalar representations of corrections at four and five points~\cite{Carrasco2020,CarrascoRodinaZekioglu2021}, and effective-theory numerator and kernel deformations~\cite{BonnefoyEtAl2022}. Kinematic Hopf algebras also provide constructions of BCJ numerators~\cite{Brandhuber2022,Chen2024a,Chen2024b}. In the construction below, we deform scalar amplitudes while keeping the field-theory kernel $S_n$ fixed.

\subsection{Disk and sphere integral amplitudes}

The normalized genus-zero disk and sphere integrals define two scalar amplitude families, $Z_n(a;\ee)$ and $J_n(x;\ee)$, with real scale parameters $a,x$~\cite{SchlottererSchnetz2018,MafraSchlotterer2016}. Their definitions and normalization are given in appendix~\ref{app:integrals}. Both families have leading term $m_n$ and three-point amplitudes equal to $m_3$ in the ordering convention of section~\ref{sec:scalarbasis}. At each order, the amplitudes have only simple massless poles in mutually compatible channels, with residues given by products of lower-point integral amplitudes~\cite{BrownDupont2019}.

Disk integrals provide the scalar coefficient functions in the complete superstring/SYM amplitude representation~\cite{MSS2011I,MSS2011II}; their low-energy expansion and pole reduction are described in ref.~\cite{BSS2013}.\footnote{Related double-copy representations express bosonic and heterotic string amplitudes in terms of scalar disk or sphere integrals and higher-derivative gauge-theory amplitudes~\cite{Azevedo2018}.}

For $Z_n(\alpha\mid\beta)$, the first ordering specifies the disk integration chamber and the second a Parke-Taylor factor, the cyclic rational function of puncture differences associated with that ordering. Field-theory KK and BCJ relations hold in the second ordering argument. For $J_n(\alpha\mid\beta)$, both orderings specify Parke-Taylor factors and both ordering arguments obey these relations~\cite{SchlottererSchnetz2018,MafraSchlotterer2016}. The sphere integrals can therefore be used in repeated scalar KLT compositions. The first disk ordering becomes the external planar ordering of the gauge amplitudes.

At four points the normalized disk factor is
\begin{align}
 Z_4(a;\ee)S_4&=B_Z(a;s,u;\ee),\notag\\
 B_Z(a;s,u;\ee)
 &=\frac{\Gamma(1-a\ee s)\Gamma(1-a\ee u)}
         {\Gamma(1-a\ee(s+u))},
 \label{eq:diskbeta}\\
 \log B_Z(a;s,u;\ee)
 &=\sum_{k\ge2}\frac{\zeta(k)a^k\ee^k}{k}P_k(s,u).
 \label{eq:disklog}
\end{align}
Choose
\begin{equation}
 a=\sqrt{\frac{c_2}{\zeta(2)}}>0.
 \label{eq:scalechoice}
\end{equation}
Euler's even-zeta identity then gives
\begin{equation}
 a^{2j}\zeta(2j)
 =-\frac{B_{2j}}{2(2j)!}(-24c_2)^j=c_{2j}.
 \label{eq:evenmatch}
\end{equation}
The disk factor already reproduces the entire even sequence. Its difference from the desired four-point logarithm is
\begin{equation}
 \log B(\ee s,\ee u)-\log B_Z(a;s,u;\ee)
 =\sum_{\substack{w\ge3\\w\ \mathrm{odd}}}
 \frac{c_w-a^w\zeta(w)}{w}\,\ee^wP_w(s,u).
 \label{eq:requireddeformation}
\end{equation}
Thus the remaining task is to adjust the odd coefficients independently while preserving higher-point factorization.

The sphere factor has the four-point expansion
\begin{equation}
 \log\!\bigl[J_4(x;\ee)S_4\bigr]
 =2\sum_{\substack{w\ge3\\w\ \mathrm{odd}}}
 \frac{\zeta(w)x^w\ee^w}{w}P_w(s,u).
 \label{eq:spherelogfour}
\end{equation}
It contains every required odd degree with a nonzero coefficient. Section~\ref{sec:construction} separates these degrees in the full sphere family and uses them to construct an auxiliary scalar family $Q_n$, with leading term $m_n$, satisfying KK and BCJ relations in both ordering arguments.

Let $\mathbf a_n^{\rm SYM}$ be the column of SYM amplitudes in the first basis of eq.~\eqref{eq:bases}, with $g^{n-2}$ removed. The gauge amplitudes will take the form
\begin{equation}
 \A_n(\alpha;\ee)
 =g^{n-2}Z_n(\alpha;a;\ee)S_nR_{Q,n}\mathbf a_n^{\rm SYM},
 \qquad R_{Q,n}=Q_nS_n.
 \label{eq:amplitudeform}
\end{equation}
Here $Z_n(\alpha;a;\ee)$ is the row with entries $Z_n(\alpha\mid\beta;a;\ee)$ as $\beta$ runs over the complementary basis. The disk integral carries the external ordering $\alpha$. The matrix $R_{Q,n}$ determines the deformation, while the dependence on the external states is entirely contained in the SYM amplitudes. Field-theory KK and BCJ relations are required for the contracted ordering indices; they are not imposed on the final deformed gauge amplitudes in their external ordering.

\subsection{Factorization of the KLT bilinear}\label{sec:factorization}

Let $X,Y$ be families of tree amplitudes defined as formal derivative expansions. Assume locality, field-theory KK and BCJ relations for the contracted ordering indices, and factorization into the corresponding lower-point amplitudes with fixed sums over intermediate states. At each derivative order, their scalar coefficients are rational functions whose denominator factors are massless channel invariants. For amplitudes that depend on external states, we work with local Lorentz tensors before choosing polarization vectors and external wavefunctions. We show that the KLT bilinear factorizes into the corresponding lower-point KLT bilinears, with the tensor product of the intermediate-state spaces, and retains this local pole structure.

Fix a channel $q=s_I$. A fixed complementary basis need not display that channel: for example, $m_4(1234\mid1243)=-1/s_{12}$ has no $s_{23}$ pole, although the ordered SYM amplitude has both adjacent poles. We first identify the basis-independent rational pairing away from $q=0$, then compute its residue in complementary bases with one fixed label in $I$ and the two final fixed labels outside $I$. Appendix~\ref{app:bases} gives the basis transformations and the rank argument for these channel-adapted bases.

The residue embeddings follow from lower-point ordering relations. An $n$-point ordering $\alpha$ compatible with the cut induces orderings $\alpha_L,\alpha_R$, each including the internal leg. Choose independent lower-point bases labeled by $\rho$ and $\sigma$. For the contracted ordering index of $Y$, write
\begin{equation}
 \begin{aligned}
 Y_L(\alpha_L;h)&=\sum_\rho C_L(\alpha_L\mid\rho)Y_L(\rho;h),\\
 Y_R(\alpha_R;\bar h)&=\sum_\sigma C_R(\alpha_R\mid\sigma)Y_R(\sigma;\bar h).
 \end{aligned}
 \label{eq:lowerorderingreductions}
\end{equation}
The KK/BCJ reduction coefficients $C_L,C_R$ are rational kinematic functions, independent of the amplitude family and of its external states. All other indices of $Y$ are held fixed in these relations. Let $\varepsilon_I(\alpha)$ be the sign fixed by the cyclic-ordering conventions in this factorization channel. Define
\begin{equation}
 U_{\alpha,(\rho,\sigma)}
 =\varepsilon_I(\alpha)\,
 C_L(\alpha_L\mid\rho)C_R(\alpha_R\mid\sigma),
 \label{eq:residueembedding}
\end{equation}
with a zero row when $I$ is incompatible with $\alpha$. Factorization then gives
\begin{equation}
 \Res_{q=0}Y_n(\alpha)
 =\sum_{\rho,\sigma}U_{\alpha,(\rho,\sigma)}\,y_{\rho\sigma},
 \qquad
 y_{\rho\sigma}=\sum_hY_L(\rho;h)Y_R(\sigma;\bar h).
 \label{eq:universalinputresidue}
\end{equation}
Reducing the lower orderings in the other $n$-point basis defines $V$ in the same way and gives $\Res_{q=0}X_n=xV^{\mathsf T}$. Here $x$ denotes the corresponding products of lower-point amplitudes, summed over the intermediate states of $X$. The signs in $U,V$ follow the same three-point cyclic-ordering conventions as $m$.

Applying these reductions to both ordering arguments of the biadjoint amplitude yields
\begin{equation}
 m_n(q)=\frac{\mathcal R}{q}+m_0+O(q),\qquad
 \mathcal R=U M V^{\mathsf T},\qquad
 M=m_L\otimes m_R.
 \label{eq:cutmatrix}
\end{equation}
Every common planar tree containing the cut is obtained by joining two lower trees. The same embeddings therefore describe the residues of $m$, $X$ and $Y$. At a generic point of $q=0$, the lower-point KK/BCJ reduction coefficients are regular and $U,V$ have full column rank
\begin{equation}
 r=(|I|-2)!\,(n-|I|-2)!.
 \label{eq:residuerank}
\end{equation}
The lower-point matrices in $M$ are invertible on their independent ordering spaces. The full-rank submatrix establishing these properties is constructed in appendix~\ref{app:bases}.

Since the $n$-point KLT kernel is polynomial, write
\begin{equation}
 S_n(q)=S_0+qS_1+O(q^2).
 \label{eq:kernelexpansion}
\end{equation}
The coefficients of $q^{-1}$ in $mS=Sm=\id$ give $\mathcal R S_0=S_0\mathcal R=0$. Full rank of $U,V$ then implies
\begin{equation}
 V^{\mathsf T}S_0=0,\qquad S_0U=0.
 \label{eq:annihilations}
\end{equation}
At order $q^0$, $\mathcal R S_1+m_0S_0=\id$. Multiplying on the right by $\mathcal R$ gives $\mathcal R S_1\mathcal R=\mathcal R$, and hence
\begin{equation}
 V^{\mathsf T}S_1U=M^{-1}.
 \label{eq:cutinverse}
\end{equation}

The residue reductions above give the complete Laurent expansions
\begin{equation}
 X_n=\frac{xV^{\mathsf T}}q+X_0+O(q),\qquad
 Y_n=\frac{Uy}q+Y_0+O(q).
 \label{eq:inputcuts}
\end{equation}
The coefficient of $q^{-2}$ in $X_nS_nY_n$ vanishes by eq.~\eqref{eq:annihilations}. At order $q^{-1}$,
\begin{align}
 \Res_{q=0}(X_nS_nY_n)
 &=xV^{\mathsf T}S_1Uy+xV^{\mathsf T}S_0Y_0+X_0S_0Uy\notag\\
 &=xM^{-1}y.
 \label{eq:residuecomposition}
\end{align}
The terms involving the regular parts $X_0$ or $Y_0$ vanish. Since $M^{-1}=S_L\otimes S_R$, the residue is the product of the lower-point KLT bilinears. When the amplitudes depend on external states, the sum runs over the tensor product of their intermediate-state spaces. A scalar factor contributes one polarization state, with its three-point amplitudes and state normalization unchanged.

If a channel is incompatible with an uncontracted ordering of $X$, then $x=0$. Its only possible contribution to the residue is $X_0S_0Uy$, which vanishes. The same statement holds for an incompatible ordering of $Y$. Applying the calculation recursively to compatible cuts gives their iterated residues. An incompatible simultaneous pole would induce a forbidden pole in a lower residue, and is therefore absent.

The polynomial kernel introduces no new denominator, and eq.~\eqref{eq:residuecomposition} removes double poles and fixes every simple residue. At each derivative order, the lower local vertices construct exchange diagrams with precisely these residues. Subtracting those diagrams cancels every possible denominator factor in the specified rational coefficient space and leaves a polynomial contact term. KLT composition therefore preserves the stated locality and factorization conditions, including the allowed channels in each uncontracted ordering and compatible iterated residues, with lower amplitudes belonging to the same composed family.

\section{All-multiplicity construction}\label{sec:construction}

To generate the four-point deformation in eq.~\eqref{eq:requireddeformation}, define the normalized matrix of sphere integrals $R_{J,n}(x;\ee)=J_n(x;\ee)S_n$ and expand its matrix logarithm:
\begin{equation}
 \log\!\bigl[J_n(x;\ee)S_n\bigr]
 =\sum_{w\ge1}x^w\ee^wL_{n,w}.
 \label{eq:Ldefinition}
\end{equation}
Here $L_{n,w}$ has degree $w$ in Mandelstam invariants. We will use the scalar families
\begin{equation}
 G_w(r)_n=\exp(r\ee^wL_{n,w})m_n,
 \qquad r\in\mathbb R.
 \label{eq:homogeneousfamily}
\end{equation}
Their four-point amplitudes follow from eq.~\eqref{eq:spherelogfour}:
\begin{equation}
 L_{4,w}=\frac{2\zeta(w)}wP_w(s,u),\qquad
 \log\!\bigl[G_w(r)_4S_4\bigr]
 =\frac{2r\zeta(w)}w\ee^wP_w(s,u),
 \quad w\ge3\ \text{odd}.
 \label{eq:generatorfour}
\end{equation}
Each family changes one odd coefficient of the four-point logarithm. Its exponential determines the higher-order terms required by factorization and retains all powers of the real parameter $r$. We establish locality and factorization of eq.~\eqref{eq:homogeneousfamily} at every multiplicity using formal matrix powers and homogeneous projection.

\subsection{Real powers of formal matrix series}\label{sec:powers}

Let $X$ be a family of doubly ordered scalar amplitudes with leading term $m$ and $X_3=m_3$ in the convention of section~\ref{sec:scalarbasis}. Assume locality, factorization through a fixed one-dimensional polarization space, and field-theory KK and BCJ relations in both ordering arguments. The normalized matrices satisfy $R_{X,3}=1$ and $R_{X,n}=\id+O(\ee)$. After truncation at Mandelstam degree $K$, define
\begin{equation}
 R_{X,n}^{\,r}
 =\sum_{j=0}^{K}\binom rj(R_{X,n}-\id)^j
 \pmod{\ee^{K+1}}.
 \label{eq:binomialpower}
\end{equation}
This finite binomial expansion is compatible with truncation and defines the power as a formal series.

For every nonnegative integer $r$, eq.~\eqref{eq:binomialpower} is repeated KLT composition. Thus $R_{X,n}^{\,r}m_n$ satisfies the required amplitude identities. At fixed multiplicity and derivative order, every coefficient is polynomial in $r$. Ordering relations and locality are linear conditions on these coefficients. Factorization is a polynomial identity involving the lower-point amplitudes at the same value of $r$. Each identity holds for every nonnegative integer and therefore holds identically in $r$. Appendix~\ref{app:formal} gives the coefficient-space formulation of this argument.

Consequently,
\begin{equation}
 X_n^{\star r}=R_{X,n}^{\,r}m_n
 \label{eq:starpower}
\end{equation}
defines scalar amplitudes satisfying locality and factorization for every real $r$. The value $r=-1$ gives the inverse under composition. At three points $R_{X,3}=1$, so $X_3^{\star r}=m_3$. The intermediate scalar polarization space and its unit normalization remain fixed.

\subsection{Homogeneous components of the matrix logarithm}\label{sec:projection}

Selecting a homogeneous term of an amplitude does not generally preserve its nonlinear factorization equations. We instead exponentiate homogeneous components of the matrix logarithm of the normalized amplitude matrix.

Consider a finite collection of scalar families $X_i$ satisfying the hypotheses of section~\ref{sec:powers}, with normalized matrices $R_{i,n}=X_{i,n}S_n$, and real coefficients $b_i$. With a fixed ordering of factors, define
\begin{equation}
 P_n(h)=\prod_i\exp\!\bigl(hb_i\log R_{i,n}\bigr).
 \label{eq:auxiliaryh}
\end{equation}
For real $h\ne0$, composition and eq.~\eqref{eq:starpower} show that $P_n(h)^{1/h}m_n$ satisfies the scalar amplitude conditions. At any fixed order in $\ee$,
\begin{equation}
 P_n(h)=\id+h\sum_i b_i\log R_{i,n}+O(h^2),\qquad
 P_n(h)^{1/h}=\exp\!\left[\frac{\log P_n(h)}h\right].
 \label{eq:regularh}
\end{equation}
Each coefficient of $\log P_n(h)$ is divisible by $h$. The second expression is therefore regular at $h=0$, where it equals
\begin{equation}
 \left.P_n(h)^{1/h}\right|_{h=0}
 =\exp\!\left(\sum_i b_i\log R_{i,n}\right).
 \label{eq:logarithmcombination}
\end{equation}
The amplitude identities are polynomial at fixed derivative order and remain satisfied at $h=0$. Exponentials of finite real linear combinations of these matrix logarithms therefore preserve locality and factorization, including when the logarithms do not commute.

Apply this operation to the sphere family at distinct nonzero real scales $x_1,\ldots,x_K$. For $1\le w\le K$, the Vandermonde system
\begin{equation}
 \sum_{j=1}^{K}b_jx_j^k=\delta_{kw},
 \qquad 1\le k\le K,
 \label{eq:vandermonde}
\end{equation}
has a unique solution. Equations~\eqref{eq:Ldefinition} and \eqref{eq:vandermonde} imply
\begin{equation}
 \sum_{j=1}^{K}b_j\log\!\bigl[J_n(x_j;\ee)S_n\bigr]
 =\ee^wL_{n,w}\pmod{\ee^{K+1}}.
 \label{eq:projectedlog}
\end{equation}
Using the coefficients $rb_j$ in eq.~\eqref{eq:logarithmcombination} proves that $G_w(r)$ in eq.~\eqref{eq:homogeneousfamily} satisfies locality and factorization through every finite order in $\ee$.

The projection coefficients depend on $w,K$ and the scales $x_j$, but not on multiplicity. They therefore act identically on the $n$-point and lower-point amplitudes in each factorization equation. Moreover, eq.~\eqref{eq:homogeneousfamily} itself is independent of $K$. Its truncations agree, so the finite-order identities establish one formal family at every order and multiplicity.

A change of field-theory ordering basis acts on $R_{X,n}$ by conjugation with a degree-zero matrix. Logarithms and homogeneous projection transform covariantly, as detailed in appendix~\ref{app:bases}. The projection uses degree in Mandelstam invariants; its validity is independent of any grading or algebraic-independence property of numerical zeta values.

Consider a six-point factorization channel with $|I|=3$, where both lower-point amplitudes have four legs. Choose two orderings compatible with the cut and suppress their indices. Let $\ell_L$ and $\ell_R$ be the four-point logarithmic coefficients evaluated on the two sides,
\begin{equation}
 \ell_L=L_{4,3}(s_L,u_L),\qquad
 \ell_R=L_{4,3}(s_R,u_R),\qquad
 L_{4,3}(s,u)=-2\zeta(3)su(s+u),
 \label{eq:sixpointgenerators}
\end{equation}
where $(s_L,u_L)$ and $(s_R,u_R)$ are their adjacent four-point invariants. Applying the factorization property just established to the auxiliary scalar family $G_3(r)$ gives
\begin{equation}
 \Res_{q=0}G_3(r)_6
 =e^{r\ee^3(\ell_L+\ell_R)}\Res_{q=0}m_6.
 \label{eq:sixpointfactorization}
\end{equation}
The signs fixed by the cyclic-ordering conventions are the same on both sides and are included in $\Res m_6$. Writing $[r^2\ee^6]$ for coefficient extraction,
\begin{equation}
 [r^2\ee^6]\Res_{q=0}G_3(r)_6
 =\left(\frac{\ell_L^2}{2}+\ell_L\ell_R+
 \frac{\ell_R^2}{2}\right)\Res_{q=0}m_6.
 \label{eq:sixpointquadratic}
\end{equation}
The cross term is the product of the first-order deformations of the two four-point factors. The other terms are their second-order exponential contributions. Keeping only $m_n+r\ee^3L_{n,3}m_n$ as a full family gives no quadratic term on the left while retaining the nonzero cross term on the right. The exponential provides the higher-order terms required by nonlinear factorization.

\subsection{Arbitrary four-point coefficients}

The disk scale $a$ in eq.~\eqref{eq:scalechoice} fixes the even coefficients. For each odd $w\ge3$, choose
\begin{equation}
 \theta_w=\frac{c_w-a^w\zeta(w)}{2\zeta(w)}.
 \label{eq:thetachoice}
\end{equation}
Since $\zeta(w)>0$, this defines a real parameter for every real $c_w$. Compose the homogeneous families in increasing order of $w$:
\begin{equation}
 Q=G_3(\theta_3)\star G_5(\theta_5)\star G_7(\theta_7)\star\cdots,
 \qquad
 R_{Q,n}=\prod_{\substack{w\ge3\\w\ \mathrm{odd}}}^{\text{increasing }w}
 \exp(\theta_w\ee^wL_{n,w}).
 \label{eq:Qmatrix}
\end{equation}
The order specifies a choice of higher-point completion, since the matrices need not commute for $n>4$. At degree $K$, only $w\le K$ contributes, and each exponential contributes finitely many terms. Hence every coefficient is defined for an arbitrary real input sequence.

Inserting eq.~\eqref{eq:Qmatrix} into eq.~\eqref{eq:amplitudeform} gives the explicit amplitudes
\begin{equation}
 \A_n(\alpha;\ee)=g^{n-2}Z_n(\alpha;a;\ee)S_n
 \left[\prod_{\substack{w\ge3\\w\ \mathrm{odd}}}^{\text{increasing }w}
 \exp(\theta_w\ee^wL_{n,w})\right]\mathbf a_n^{\rm SYM}.
 \label{eq:finalconstruction}
\end{equation}
In the reference three-point basis, the scalar amplitudes and $S_3$ equal one, so $\A_3=\A_3^{\rm SYM}$. At four points the normalized matrices are scalars and their logarithms add:
\begin{align}
 \log\frac{\A_4}{\A_4^{\rm SYM}}
 &=\sum_{j\ge1}\frac{c_{2j}\ee^{2j}}{2j}P_{2j}
 +\sum_{\substack{w\ge3\\w\ \mathrm{odd}}}
 \frac{a^w\zeta(w)+2\theta_w\zeta(w)}{w}\ee^wP_w\notag\\
 &=\sum_{k\ge2}\frac{c_k\ee^k}{k}P_k.
 \label{eq:fourmatch}
\end{align}
Thus $\A_4=B(\ee s,\ee u)\A_4^{\rm SYM}$, which is eq.~\eqref{eq:result} with its $\ee$ dependence displayed. The operations defining the scalar amplitudes in eq.~\eqref{eq:finalconstruction} preserve locality and factorization. The remaining physical properties follow from the SYM amplitudes.

\subsection{The physical multiplet and symmetries}\label{sec:physical}

The SYM on-shell superfield~\cite{EFK2010} is
\begin{equation}
 \Phi(\eta)=g^++\eta^A\psi_A
 +\frac12\eta^A\eta^B\phi_{AB}
 +\frac1{3!}\epsilon_{ABCD}\eta^A\eta^B\eta^C\bar\psi^D
 +\eta^1\eta^2\eta^3\eta^4g^-,
 \label{eq:superfield}
\end{equation}
where $A=1,\ldots,4$ and the scalars satisfy the standard reality condition. It contains both gluon helicities, four fermions, six real scalars, and four antifermions. Each auxiliary scalar factor in eq.~\eqref{eq:amplitudeform} contributes a one-dimensional polarization space. Equation~\eqref{eq:residuecomposition} therefore gives
\begin{equation}
 \Res_{s_I=0}\A_n
 =\int d^4\eta_I\,
 \A_L(\ldots,P_I,\eta_I)\,
 \A_R(-P_I,\eta_I,\ldots),
 \label{eq:supersymmetriccut}
\end{equation}
with the standard crossed-momentum phases. This is the complete SYM physical-state sum with its original positive inner product.

All coefficients multiplying the SYM amplitudes are functions of the Mandelstam invariants independent of $\eta$. They commute with
\begin{equation}
 Q^A_\alpha=\sum_i\lambda_{i\alpha}\eta_i^A,\qquad
 \widetilde Q_{A\dot\alpha}
 =\sum_i\tilde\lambda_{i\dot\alpha}\frac{\partial}{\partial\eta_i^A},
 \label{eq:supercharges}
\end{equation}
and preserve the Lorentz, gauge, and $SU(4)$ Ward identities. They also commute with the Grassmann integration in eq.~\eqref{eq:supersymmetriccut}.

For a SYM tree amplitude with only external scalars, an internal fermion subgraph would have two fermion lines at every vertex and no external fermionic endpoint. A nonempty such subgraph contains a cycle and therefore cannot occur in a tree. The remaining bosonic gauge and scalar vertices contain momenta, the spacetime metric, and scalar internal tensors, with no four-dimensional Levi-Civita contraction. The scalar SYM amplitudes consequently satisfy eq.~\eqref{eq:scalarparity}. Linear combinations with coefficients depending on the Mandelstam invariants preserve this condition with all scalar labels fixed.

Cyclicity and reflection in the external ordering follow from the disk amplitudes. The integral identities and KLT basis invariance give covariance under relabeling.

All coefficients and parameters are real. The scalar rational identities hold at generic Mandelstam invariants satisfying momentum conservation and extend to four-dimensional kinematics by continuation; the construction introduces no inverse Gram determinant.

The leading scalar degree $-(n-3)$ cancels the kernel degree $n-3$, so the transformation is dimensionless at leading order. In physical units, each correction of Mandelstam degree $w$ is accompanied by the factor $\Lambda^{-2w}$ relative to SYM. Since $n_L+n_R=n+2$, the coupling powers satisfy $(n_L-2)+(n_R-2)=n-2$ on every cut. Equation~\eqref{eq:finalconstruction} therefore has the spectrum, symmetries, normalization, and factorization required in section~\ref{sec:data}.

The amplitudes also inherit the disk monodromy relations in the ordering that specifies the integration domain~\cite{Plahte1970,BBDV2009,Stieberger2009}. For the doubly ordered disk integrals, they are reviewed in ref.~\cite{MafraSchlotterer2016}. Transporting a boundary puncture past the others gives linear relations with coefficients determined by the Koba-Nielsen exponents. In a common contracted basis, let $C_\alpha(s;a\ee)$ denote the coefficients of any such relation, so that
\begin{equation}
 \sum_\alpha C_\alpha(s;a\ee)\,
 Z_n(\alpha\mid\beta;a;\ee)=0
 \qquad\text{for every }\beta.
 \label{eq:diskmonodromy}
\end{equation}
Here $s$ denotes the full set of kinematic invariants and the sum runs over the orderings in the relation. Every factor to the right of the disk row in eq.~\eqref{eq:finalconstruction} is independent of $\alpha$. Therefore
\begin{align}
 \sum_\alpha C_\alpha(s;a\ee)\,\A_n(\alpha;\ee)
 &=g^{n-2}\left[\sum_\alpha C_\alpha(s;a\ee)\,
 Z_n(\alpha;a;\ee)\right]
 S_nR_{Q,n}\mathbf a_n^{\rm SYM}\notag\\
 &=0.
 \label{eq:inheritedmonodromy}
\end{align}
These identities hold coefficientwise in $\ee$. The constructed completions thus obey disk monodromy at the scale $a$ fixed by $c_2$. This is an additional property of the chosen representatives; the existence statement imposes only the amplitude conditions in section~\ref{sec:data}.

\section{Examples and higher-point nonuniqueness}\label{sec:examples}

\subsection{The open-superstring amplitude and the case of vanishing odd coefficients}

For
\begin{equation}
 c_k=a^k\zeta(k),\qquad k\ge2,
 \label{eq:stringchoice}
\end{equation}
every $\theta_w$ vanishes. Then $Q=m$, and eq.~\eqref{eq:amplitudeform} reduces to the disk-integral representation of massless open-superstring tree amplitudes expanded about the SYM limit~\cite{MSS2011I,MSS2011II}.

A second example sets every odd coefficient to zero. One completion follows directly from the formal power
\begin{equation}
 Q_n=\bigl[J_n(a;\ee)S_n\bigr]^{-1/2}m_n.
 \label{eq:evencompletion}
\end{equation}
Its four-point logarithm cancels the odd part of the disk logarithm. With the scale $\kappa$ defined in section~\ref{sec:data}, so that $\kappa=\pi a$, and the $\ee$ dependence suppressed, eq.~\eqref{eq:evensolution} gives the multiplier
\begin{equation}
 B_0(s,u)=\left[
 \frac{(\kappa s/\sin\kappa s)(\kappa u/\sin\kappa u)}
 {\kappa(s+u)/\sin\!\bigl(\kappa(s+u)\bigr)}
 \right]^{1/2}.
 \label{eq:evenexample}
\end{equation}
The square root is the formal expansion with constant term one at the origin. Its generic analytic continuation has square-root singularities, in particular at nonzero zeros of $\sin(\kappa s)$ when the remaining factors are finite and nonzero. Equation~\eqref{eq:evencompletion}, inserted into eq.~\eqref{eq:amplitudeform}, defines the amplitudes as formal series at all multiplicities. It is an alternative to the product ordered by increasing Mandelstam degree in eq.~\eqref{eq:Qmatrix}: the two prescriptions have identical four-point data, while their higher-point matrices can differ.

\subsection{A five-point contact deformation}

The commutator of the string coefficient matrices multiplying $\zeta(3)$ and $\zeta(5)$ already occurs at degree eight in the low-energy expansion~\cite{SchlottererStieberger2012}; the corresponding higher-point freedom was also discussed in ref.~\cite{EHM2026}. We use this kinematic structure to construct a real one-parameter deformation that preserves the entire four-point amplitude and satisfies locality and factorization at every multiplicity. Define the scalar group commutator
\begin{equation}
 K(r)=G_3(r)\star G_5(1)\star G_3(-r)\star G_5(-1),
 \qquad r\in\mathbb R.
 \label{eq:groupcommutator}
\end{equation}
Composition and formal inverses preserve its locality and factorization. At three points $K_3(r)=m_3$, while at four points all normalized matrices commute, giving
\begin{equation}
 K_4(r)=m_4.
 \label{eq:fourtrivial}
\end{equation}
At higher multiplicity,
\begin{equation}
 R_{K,n}=\id+r\ee^8[L_{n,3},L_{n,5}]+O(\ee^{11}).
 \label{eq:commutatorexpansion}
\end{equation}
Replace $Q$ in eq.~\eqref{eq:amplitudeform} by $K(r)\star Q$, and denote the resulting amplitudes by $\A_n^{(r)}$. The disk row is unchanged, so these amplitudes obey the same monodromy relations~\eqref{eq:inheritedmonodromy}. Their difference $\Delta\A_n=\A_n^{(r)}-\A_n$ obeys
\begin{equation}
 \Delta\A_3=\Delta\A_4=0,\qquad
 \Res_{s_I=0}\Delta\A_5=0.
 \label{eq:contactresidues}
\end{equation}
The second equality follows because every five-point factorization channel is a product of a three-point and a four-point amplitude, both unchanged. Locality then implies that $\Delta\A_5$ is a contact amplitude at each derivative order. It preserves supersymmetry, $SU(4)$, and the scalar parity condition. Equation~\eqref{eq:groupcommutator} defines its extension to all higher multiplicities, including the corresponding factorization contributions.

The deformation is nonzero. Appendix~\ref{app:checks} gives exact momentum-conserving four-dimensional spinors and evaluates the five-point matrices from the standard disk recursion. The commutator $[L_{5,3},L_{5,5}]$ acts nontrivially on the SYM maximally helicity-violating (MHV) amplitude vector. Since the other normalized factors have leading term $\id$, the resulting degree-eight contact coefficient is nonzero for $r\ne0$, independently of the fixed four-point data.\footnote{Degree eight adds sixteen powers of momenta to the five-point SYM scaling $p^{-1}$. The five-gluon contact term therefore scales as $p^{15}$ and corresponds by dimensional counting to dimension-20 operators, schematically $D^{10}F^5$, where $D$ is a gauge-covariant derivative and $F_{\mu\nu}$ the field strength.} Thus every member of the four-point family has inequivalent higher-point completions with the same disk monodromy relations.

\section{Discussion}\label{sec:discussion}

Every four-point function in the maximally supersymmetric $c_2>0$ class of section~\ref{sec:data} admits planar massless tree amplitudes at all multiplicities and derivative orders. KLT composition preserves factorization. Exponentiating homogeneous components of the matrix logarithm $\log R_{J,n}$ varies the odd four-point coefficients independently. Only finitely many factors contribute at each derivative order, so the construction applies to every real sequence and gives lower-point amplitudes satisfying every factorization equation. Combined with the necessary six-point identities~\eqref{eq:cocycle} and \eqref{eq:sixpointidentity}, this establishes the necessary-and-sufficient conditions for the four-point amplitudes to admit completion.

Higher-point consistency constrains the accompanying contact interactions, but these are not uniquely fixed by the four-point amplitude. The normalized matrices are scalars at four points and need not commute at higher multiplicity. The group commutator constructed in section~\ref{sec:examples} gives a nonzero five-point contact amplitude and a consistent higher-point extension. It preserves the entire four-point function and the same disk monodromy relations. Thus the four-point conditions are complete, while the higher-point amplitudes are not unique.

Additional spectral and dispersive assumptions can select a smaller family. In ref.~\cite{Sung2026}, meromorphy, a gapped massive spectrum with nonnegative scalar partial-wave residues, and an exact unsubtracted forward dispersion relation select the Veneziano amplitude. Monodromy and dispersive positivity provide complementary restrictions~\cite{HuangEtAl2021,BermanEtAl2024,ChiangHuangWeng2024}. Analytic reconstruction with spectral assumptions and either monodromy or splitting is studied in ref.~\cite{WanZhou2026}. Multiparticle factorization also excludes specified deformations when their massive exchange content is fixed~\cite{ArkaniHamedEtAl2024,BasileRemmenStaudt2026}. The present result determines the allowed four-point amplitudes in the formal massless derivative expansion, with higher-point contact interactions satisfying the stated symmetries and factorization conditions. Within this amplitude class, the four-point constraints are complete at six points.

\acknowledgments
The author thanks Ryumi S. for inspiration and Damian G. and Chris M. for useful discussions, and Yu-tin Huang for valuable insights. The views expressed are those of the author and do not necessarily represent the U.S. Department of Homeland Security or the United States government.

\appendix

\section{Disk and sphere integrals}\label{app:integrals}

\subsection{Measures and normalization}

For $z_{ij}=z_i-z_j$, define the Parke-Taylor factor
\begin{equation}
 \PT(\beta)=\frac1{z_{\beta_1\beta_2}z_{\beta_2\beta_3}\cdots z_{\beta_n\beta_1}}.
 \label{eq:PTdefinition}
\end{equation}
Let $\nu_{ij}=\nu_{ji}$, $\nu_{ii}=0$, and $\sum_j\nu_{ij}=0$ be the Koba-Nielsen exponents. With the cyclic-ordering conventions of refs.~\cite{SchlottererSchnetz2018,MafraSchlotterer2016}, the disk and sphere integrals are
\begin{align}
 I^Z_n(\alpha\mid\beta;\nu)
 &=(-1)^{n-3}\int_{C_\alpha}
 \frac{\prod_i dz_i}{\operatorname{vol}SL(2,\mathbb R)}
 \prod_{i<j}|z_{ij}|^{\nu_{ij}}\,\PT(\beta),
 \label{eq:rawdisk}\\
 I^J_n(\alpha\mid\beta;\nu)
 &=\int_{\mathcal M_{0,n}(\mathbb C)}
 \frac{\prod_i d^2z_i}{\pi^{n-3}\operatorname{vol}SL(2,\mathbb C)}
 \prod_{i<j}|z_{ij}|^{2\nu_{ij}}\,
 \overline{\PT(\alpha)}\PT(\beta).
 \label{eq:rawsphere}
\end{align}
Here $C_\alpha$ is the real chamber with the punctures in cyclic order $\alpha$, and $\mathcal M_{0,n}$ is the moduli space of $n$ ordered distinct points on the sphere. Fixing three real punctures $z_i,z_j,z_k$ removes their integrations and inserts $|z_{ij}z_{ik}z_{jk}|$. The sphere measure contains the square of this factor, with $d^2z=d(\Re z)d(\Im z)$. The signs follow from the cyclic-ordering conventions of section~\ref{sec:scalarbasis}.

Write $d=n-3$. The normalized families in the propagator convention of section~\ref{sec:data} are
\begin{equation}
 Z_n(a;\ee)=(-a\ee)^d I^Z_n(-a\ee s),\qquad
 J_n(x;\ee)=(-x\ee)^d I^J_n(-x\ee s).
 \label{eq:normalizedintegrals}
\end{equation}
The notation on the right denotes substitution for every exponent. The factors remove the common leading degree $-d$, leaving $m_n(s)$ as the leading term. At zero scale the expression is defined by its formal expansion. The three-point integral amplitudes equal $m_3$ in the same ordering convention.

The complementary four-point entry fixes the relative signs. With $(z_1,z_3,z_4)=(0,1,\infty)$, $v=\nu_{12}$, $w=\nu_{23}$, and $z=z_2$,
\begin{equation}
 I^Z_4(1234\mid1243)
 =-\int_0^1\frac{dz}{z}\,z^v(1-z)^w
 =-\frac1v\frac{\Gamma(1+v)\Gamma(1+w)}{\Gamma(1+v+w)}.
 \label{eq:rawfour}
\end{equation}
The normalized entry is $-B_Z/s$, so $m_4=-1/s$ and $S_4=-s$. The minus sign follows from reversal of a cyclic ordering in a three-point amplitude; the propagator remains $+1/s$. Multiplying the normalized disk entry by $S_4$ gives eq.~\eqref{eq:diskbeta}.

The corresponding complex beta integral is
\begin{equation}
 \frac1\pi\int_{\mathbb C}
 \frac{d^2z}{z\bar z(1-\bar z)}|z|^{2v}|1-z|^{2w}
 =\frac{\Gamma(v)\Gamma(1+w)\Gamma(1-v-w)}
        {\Gamma(1-v)\Gamma(1-w)\Gamma(1+v+w)}.
 \label{eq:complexbeta}
\end{equation}
The complementary sphere entry has the same overall minus sign as eq.~\eqref{eq:rawfour}. Normalization and multiplication by $S_4$ give
\begin{equation}
 J_4(x;\ee)S_4
 =\frac{\Gamma(1-x\ee s)\Gamma(1-x\ee u)\Gamma(1+x\ee(s+u))}
        {\Gamma(1+x\ee s)\Gamma(1+x\ee u)\Gamma(1-x\ee(s+u))}.
 \label{eq:spherebeta}
\end{equation}
Its logarithmic expansion is eq.~\eqref{eq:spherelogfour}.

\subsection{Ordering relations and residues}

The KK relations follow from rational identities among Parke-Taylor factors. To obtain a fundamental BCJ relation, leave $z_1$ unfixed and insert label 1 into the chain of remaining labels:
\begin{equation}
 \PT(2,\ldots,j,1,j+1,\ldots,n)
 =\PT(2,\ldots,n)
 \left(\frac1{z_j-z_1}-\frac1{z_{j+1}-z_1}\right).
 \label{eq:PTinsertion}
\end{equation}
Weighting by $\nu_{12}+\cdots+\nu_{1j}$ and summing gives
\begin{equation}
 -\PT(2,\ldots,n)\,\partial_{z_1}\log\mathrm{KN},
 \label{eq:BCJtotalderivative}
\end{equation}
where $\mathrm{KN}$ is the Koba-Nielsen factor. After multiplication by $\mathrm{KN}$, this is a total derivative. Its integral vanishes in a convergence region and hence in the meromorphic continuation and canonical Laurent expansion. On the sphere, the antiholomorphic calculation gives the relation in the other ordering argument. On the disk, the first ordering specifies the integration domain, so this derivation applies only to the second ordering.

A massless channel corresponds to a boundary divisor separating a subset $I$ of punctures from its complement. The boundary geometry and its relation to KLT are developed in ref.~\cite{Mizera2017}. Let $\nu_I=\sum_{i<j\in I}\nu_{ij}$ and let $q$ be a local coordinate transverse to the boundary divisor. Integration over this transverse coordinate gives $q^{\nu_I}dq/q$ on the disk and $|q|^{2\nu_I}d^2q/(\pi|q|^2)$ on the sphere. Both give a simple pole $1/\nu_I$. The logarithmic forms restrict to products of lower-point Parke-Taylor forms, and the integration domains factorize accordingly. Incompatibility with an ordering makes the logarithmic residue vanish.

The canonical Laurent expansions and residue formulas are given by corollaries~4.21 and 4.22 of ref.~\cite{BrownDupont2019} for the disk integrals and corollaries~4.25 and 4.26 for the sphere integrals. Applied to these logarithmic Parke-Taylor forms and their factorized boundary restrictions, they yield
\begin{equation}
 \Res_{s_I=0}Z_n(\alpha\mid\beta)
 =\sigma(\alpha,\beta;I)Z_L(\alpha_L\mid\beta_L)Z_R(\alpha_R\mid\beta_R),
 \label{eq:scalarintegralcut}
\end{equation}
with the same expression for $J$. The sign $\sigma$ is fixed by the three-point cyclic-ordering conventions in factorization and is included in $U,V$ in eq.~\eqref{eq:cutmatrix}. Normalization is preserved because $d_L+d_R=d-1$: the $n$-point normalization divided by the transverse exponent gives the product of the two lower-point normalizations. Compatible boundary divisors correspond to a common polygon dissection, and their iterated residues are independent of the cut order. These residue formulas and restrictions on simultaneous poles establish the scalar factorization and locality used in section~\ref{sec:klt}.

\section{KLT bases on factorization channels}\label{app:bases}

\subsection{Basis transformations}

In this subsection, let $T,U$ denote changes of the independent left and right ordering bases. Then
\begin{equation}
 m'=TmU^{\mathsf T},\qquad
 S'=U^{-\mathsf T}ST^{-1},\qquad
 X'=TXU^{\mathsf T},\qquad
 R_X'=TR_XT^{-1}.
 \label{eq:basistransforms}
\end{equation}
For a scalar row $z$ and an amplitude column $a$, $z'=zU^{\mathsf T}$ and $a'=Ta$, so $z'S'a'=zSa$. The lower-point KLT bilinears are therefore independent of the basis, including when the inverse lower-point matrix in an intermediate basis is rational rather than polynomial. The basis-change matrices have degree zero in Mandelstam invariants. Conjugation therefore preserves the homogeneous decomposition in eq.~\eqref{eq:Ldefinition}, as well as powers and logarithms.

\subsection{Channel-adapted bases and residue rank}

For $2\le |I|\le n-2$, choose one fixed external label $i\in I$ and two fixed external labels $j,k\notin I$. The complementary $n$-point bases end in $(j,k)$ and $(k,j)$. Choose also $b\in I\setminus\{i\}$. Consider the $n$-point orderings with the contiguous block
\begin{equation}
 (i,\sigma(I\setminus\{i,b\}),b)
 \label{eq:adaptedblock}
\end{equation}
followed by a permutation of the remaining labels outside $I$. Cutting these orderings gives pairs of lower-point bases. Writing $\ell$ for the internal label, the $I$ side has the basis $(i,\sigma,b,\ell)$ in both orderings; the other side has the complementary bases $(\ell,\tau,j,k)$ and $(\ell,\tau,k,j)$. Their dimensions are $(|I|-2)!$ and $(n-|I|-2)!$.

Each common planar cubic tree containing the cut is obtained by joining two lower trees. The selected submatrix of the residue is therefore $m_L\otimes m_R$. Both lower matrices are generically invertible. For the matrix on the $I$ side, which has the same ordering basis for rows and columns, invertibility follows from KLT orthogonality: its scattering-equation representation is $W^{\mathsf T}DW$, with $W$ the invertible square matrix of basis Parke-Taylor evaluations and $D$ an invertible diagonal matrix~\cite{CHYOrthogonality2014,CHY2013}. All other orderings compatible with the channel reduce to the same lower-point bases by KK and BCJ relations. The full residue rank is consequently $(|I|-2)!\,(n-|I|-2)!$, and $U,V$ in eq.~\eqref{eq:cutmatrix} have full column rank.

The embeddings in eq.~\eqref{eq:residueembedding} are obtained from lower-point KK/BCJ reductions at generic kinematics on the factorization channel, where the lower-point basis matrices are invertible. Changes of the $n$-point basis can nevertheless be singular at $q=0$. Equation~\eqref{eq:basistransforms} identifies the rational KLT bilinears away from the divisor; the adapted representation of section~\ref{sec:factorization} then determines their common residue. No regularity of an $n$-point basis transformation on the cut is required. Repeating this prescription on lower factors gives compatible iterated residues, and the rational identities extend to four-dimensional kinematics by continuation.

\section{Formal series and homogeneous projection}\label{app:formal}

At fixed multiplicity, let $\mathcal V$ be the real vector space of rational kinematic coefficients with the required ordering indices, and let $\mathcal V_{\rm loc}$ be the linear subspace with the prescribed local pole structure. Modulo $\ee^{K+1}$, eq.~\eqref{eq:binomialpower} is a finite polynomial in $r$ with coefficients in $\mathcal V$. Its image in $\mathcal V/\mathcal V_{\rm loc}$ vanishes for every nonnegative integer $r$. A vector-valued polynomial has finitely many coefficients; evaluations at sufficiently many distinct real values invert a scalar Vandermonde matrix. Hence every coefficient of the projected polynomial vanishes. This gives locality at arbitrary real $r$, and the same argument applies to the linear ordering and symmetry identities.

For factorization, write $X_n^{\star r}[k]$ for the coefficient of $\ee^k$ after the leading scaling is removed. At each order the difference
\begin{equation}
 \Res X_n^{\star r}[k]
 -\sum_{i+j=k}X_L^{\star r}[i]X_R^{\star r}[j]
 \label{eq:polynomialfactorization}
\end{equation}
is polynomial in $r$. Products include the fixed orientation signs and state contractions. The expression vanishes at all nonnegative integers by KLT composition, so it vanishes identically. Since $R_X-\id=O(\ee)$, increasing the truncation order leaves every earlier coefficient unchanged.

The logarithmic operation has the same finite-order interpretation. At fixed truncation order, every coefficient of $P_n(h)$ in eq.~\eqref{eq:auxiliaryh} is polynomial in $h$, and
\begin{equation}
 \log P_n(h)=h\sum_i b_i\log R_{i,n}+h^2Y_n(h),
 \label{eq:loghregularity}
\end{equation}
where $Y_n(h)$ has polynomial coefficients at that order. Dividing by $h$ and exponentiating is regular at $h=0$. The amplitude identities hold for every nonzero real $h$, and their polynomial coefficients therefore retain those identities upon setting $h=0$. Noncommuting products occur in $Y_n(h)$ and do not affect the value at $h=0$.

For the homogeneous projection, the matrix $V_{kj}=x_j^k$, $1\le k,j\le K$, has determinant
\begin{equation}
 \det V=\left(\prod_{j=1}^Kx_j\right)
 \prod_{i<j}(x_j-x_i).
 \label{eq:Vdeterminant}
\end{equation}
Its nonvanishing gives eq.~\eqref{eq:projectedlog} at each truncation order. The resulting exponential in eq.~\eqref{eq:homogeneousfamily} is independent of the truncation order and uses the same projection at every multiplicity. Finally, at degree $K$ the ordered product in eq.~\eqref{eq:Qmatrix} contains only parameters with $w\le K$, each to a power at most $\lfloor K/w\rfloor$. This establishes the construction coefficient by coefficient for every real input sequence.

\section{The five-point contact deformation}\label{app:checks}

\subsection{Exact four-dimensional evaluation}

Choose dimensionless spinors
\begin{equation}
 \lambda_i=(1,i-1),\qquad i=1,\ldots,5,
 \label{eq:fivespinors}
\end{equation}
with
\begin{equation}
 \begin{gathered}
 \tilde\lambda_1=(1,0),\quad \tilde\lambda_2=(0,1),\quad
 \tilde\lambda_3=(1,1),\\
 \tilde\lambda_4=(-6,-5),\qquad \tilde\lambda_5=(4,3).
 \end{gathered}
 \label{eq:fivetildes}
\end{equation}
They satisfy $\sum_i\lambda_i\tilde\lambda_i^{\mathsf T}=0$. In the convention of eq.~\eqref{eq:spinors}, the two-particle Mandelstam invariants are
\begin{equation}
 \begin{gathered}
 (s_{12},s_{13},s_{14},s_{15})=(1,2,-15,12),\\
 (s_{23},s_{24},s_{25})=(-1,12,-12),\qquad
 (s_{34},s_{35},s_{45})=(1,-2,2).
 \end{gathered}
 \label{eq:fiveinvariants}
\end{equation}
All denominators below are nonzero.

Use the SYM ordering basis $(12345,13245)$. Let $\mathcal F_5^+(\ee s)$ be the normalized disk matrix in the sign convention in which the four-point beta-function factor contains $\Gamma(1+\ee s)$, as in eq.~\eqref{eq:KZendpoints}. Following the coefficient-matrix notation of ref.~\cite{SchlottererStieberger2012}, denote its coefficients multiplying $\ee^3\zeta(3)$ and $\ee^5\zeta(5)$ by $M_3$ and $M_5$. At the kinematics~\eqref{eq:fivespinors} and \eqref{eq:fivetildes},
\begin{equation}
 M_3=\begin{pmatrix}162&-360\\-27&384\end{pmatrix},\qquad
 M_5=\begin{pmatrix}26406&-69912\\-8847&75696\end{pmatrix}.
 \label{eq:oddmatrices}
\end{equation}
The recursion in the following subsection determines these entries. Their commutator is
\begin{equation}
 \bigl[M_3,M_5\bigr]=
 \begin{pmatrix}
 1297296&-2223936\\
 -633204&-1297296
 \end{pmatrix}.
 \label{eq:actualcommutator}
\end{equation}
For the gluon component with negative helicities on legs 1 and 2, the coupling-stripped Parke-Taylor amplitudes~\cite{ParkeTaylor1986} give
\begin{equation}
 v_{\rm MHV}=\begin{pmatrix}-1/4\\1/16\end{pmatrix},\qquad
 [M_3,M_5]v_{\rm MHV}
 =\begin{pmatrix}-463320\\77220\end{pmatrix}\ne0.
 \label{eq:physicalaction}
\end{equation}

The substitution in eq.~\eqref{eq:normalizedintegrals} multiplies a degree-$w$ coefficient by $(-1)^w$. At degrees three and five, the sphere expansion multiplies the terms proportional to $\zeta(3)$ and $\zeta(5)$ by two and removes terms containing even zeta values~\cite{BrownSingleValued2014,SchlottererSchnetz2018,BrownDupont2019}. These are low-degree instances of the single-valued map; a generating-series description at genus zero is developed in ref.~\cite{FrostEtAl2025}. The corresponding homogeneous components of $\log R_{J,5}$ are
\begin{equation}
 L_{5,3}=-2\zeta(3)M_3,\qquad
 L_{5,5}=-2\zeta(5)M_5,
 \label{eq:Lspherefive}
\end{equation}
and
\begin{equation}
 \bigl[L_{5,3},L_{5,5}\bigr]=4\zeta(3)\zeta(5)[M_3,M_5].
 \label{eq:physicalcommutator}
\end{equation}
The coefficient of $\ee^8$ in the deformation of section~\ref{sec:examples} is consequently
\begin{equation}
 \Delta\boldsymbol{\A}_5[8]
 =4r g^3\zeta(3)\zeta(5)
 \begin{pmatrix}-463320\\77220\end{pmatrix}.
 \label{eq:contactcoefficient}
\end{equation}
At this first nonzero order, all other normalized factors contribute their leading identity. The coefficient is therefore independent of the specified four-point data. Equations~\eqref{eq:contactresidues} and \eqref{eq:contactcoefficient} establish a nonzero local five-point deformation. Since it changes an on-shell amplitude, this deformation cannot be removed by a local field redefinition.

\subsection{Coefficient matrices from the KZ equation}

The five-point Knizhnik-Zamolodchikov (KZ) recursion~\cite{BSST2013} determines the matrices in eq.~\eqref{eq:oddmatrices}. In this subsection only, abbreviate
\begin{equation}
 a=s_{12},\quad b=s_{13},\quad c=s_{23},\quad
 d=s_{24},\quad e=s_{34}.
 \label{eq:KZvariables}
\end{equation}
The auxiliary six-component system obeys
\begin{equation}
 \frac{d\widehat F}{dz}
 =\ee\left(\frac{E_0}{z}+\frac{E_1}{1-z}\right)\widehat F,
 \label{eq:KZequation}
\end{equation}
where
\begin{equation}
 E_0=\begin{pmatrix}
 a+b+c&0&-b-c&-a&-a&a\\
 0&a+b+c&-b&-a-c&b&-b\\
 0&0&a&0&-a&0\\
 0&0&0&b&0&-b\\
 0&0&0&0&0&0\\
 0&0&0&0&0&0
 \end{pmatrix},
 \label{eq:KZEzero}
\end{equation}
\begin{equation}
 E_1=\begin{pmatrix}
 0&0&0&0&0&0\\
 0&0&0&0&0&0\\
 e&0&-e&0&0&0\\
 0&d&0&-d&0&0\\
 e&-e&c+d&e&-c-d-e&0\\
 -d&d&d&c+e&0&-c-d-e
 \end{pmatrix}.
 \label{eq:KZEone}
\end{equation}
The Drinfeld associator $\Phi_{\mathrm{KZ}}$ relates the regularized boundary values~\cite{DrummondRagoucy2013}:
\begin{equation}
 C_1=\Phi_{\mathrm{KZ}}(\ee E_0,\ee E_1)C_0,\qquad
 C_0=(F_4^+,0,0,0,0,0)^{\mathsf T},\qquad
 F_4^+=\frac{\Gamma(1+\ee a)\Gamma(1+\ee c)}
              {\Gamma(1+\ee(a+c))}.
 \label{eq:KZendpoints}
\end{equation}
The first two components of $C_1$ give the first row of $\mathcal F_5^+$. Exchanging labels 2 and 3 and interchanging the columns gives the second row.

Let $I(w)$ denote the shuffle-regularized iterated integral in $\omega_0=dz/z$ and $\omega_1=dz/(1-z)$, with the letters of $w=(w_1,\ldots,w_k)$ ordered from the innermost to the outermost integration. Then
\begin{equation}
 \Phi_{\mathrm{KZ}}(\ee E_0,\ee E_1)
 =\id+\sum_{k\ge1}\ee^k\sum_{w\in\{0,1\}^k}
 I(w)E_{w_k}\cdots E_{w_1}.
 \label{eq:wordassociator}
\end{equation}
\begin{table}[!htbp]
 \centering
 \begin{tabular}{lrr}
 \toprule
 Words $w$ & coefficient of $\zeta(5)$ & coefficient of $\zeta(2)\zeta(3)$\\
 \midrule
 $10000$, $11110$ & $1$ & $0$\\
 $10010$, $10110$ & $9/2$ & $-2$\\
 $10100$, $11010$ & $-11/2$ & $3$\\
 $11000$, $11100$ & $2$ & $-1$\\
 \bottomrule
 \end{tabular}
 \caption{Convergent iterated integrals of weight five in eq.~\eqref{eq:wordassociator}. Each word in a row has the indicated coefficients. The word order follows the inner-to-outer integration convention.}
 \label{tab:words}
\end{table}

The iterated integrals for words beginning in 1 and ending in 0 converge; shuffle regularization with $I(0)=I(1)=0$ defines the others. The values $I(10)=\zeta(2)$ and $I(100)=I(110)=\zeta(3)$, table~\ref{tab:words}, and the expansion of the beta-function factor in eq.~\eqref{eq:KZendpoints} determine the matrices~\eqref{eq:oddmatrices} and the nonzero contact amplitude~\eqref{eq:contactcoefficient}.

\bibliographystyle{JHEP}
\bibliography{references}
\end{document}